\documentclass[letterpaper,twocolumn,10pt]{article}
\usepackage{usenix,epsfig,endnotes}
\usepackage{mathtools}
\usepackage{amsthm}
\usepackage[boxruled,linesnumbered]{algorithm2e}
\usepackage{url}
\usepackage{relsize}
\usepackage{cite}
\usepackage[breaklinks,colorlinks=true,citecolor=blue,pdftex]{hyperref}
\usepackage{bm}
\usepackage{changepage}

\begin{document}

\title{DCCast: Efficient Point to Multipoint Transfers Across Datacenters}

\author{
    \vspace{0.5em}
    {\rm Mohammad Noormohammadpour}$^\dagger$, {\rm Cauligi S. Raghavendra}$^\dagger$, {\rm Sriram Rao}$^\ddagger$, {\rm Srikanth Kandula}$^\ddagger$\\
    $^\dagger$Ming Hsieh Department of Electrical Engineering, University of Southern California\\
    $^\ddagger$Microsoft\\
}

\maketitle

\pagestyle{empty}

\subsection*{Abstract}
Using multiple datacenters allows for higher availability, load balancing and reduced latency to customers of cloud services. To distribute multiple copies of data, cloud providers depend on inter-datacenter WANs that ought to be used efficiently considering their limited capacity and the ever-increasing data demands. In this paper, we focus on applications that transfer objects from one datacenter to several datacenters over dedicated inter-datacenter networks. We present DCCast, a centralized Point to Multi-Point (P2MP) algorithm that uses forwarding trees to efficiently deliver an object from a source datacenter to required destination datacenters. With low computational overhead, DCCast selects forwarding trees that minimize bandwidth usage and balance load across all links. With simulation experiments on Google's GScale network, we show that DCCast can reduce total bandwidth usage and tail Transfer Completion Times (TCT) by up to $50\%$ compared to delivering the same objects via independent point-to-point (P2P) transfers.

\section{Introduction}
Increasingly, companies rely on multiple datacenters to improve quality of experience for their customers. The benefits of having multiple datacenters include reduced latency to customers by mapping users to datacenters according to location, increased failure resiliency and availability, load balancing by mapping users to different datacenters and respecting local data laws. Companies may either own the datacenters or depend on infrastructure and services from providers such as Microsoft Azure \cite{azure}, Google Compute Engine \cite{google} or Amazon Web Services \cite{aws}. Large providers use geographically distributed dedicated networks to connect their datacenters \cite{orchestrating, tempus, swan, b4}.

Nowadays, many services operate across multiple datacenters. Such services require efficient data transfers among datacenters including replication of objects from one datacenter to multiple datacenters which is referred to as geo-replication \cite{mesa, mdcc, owan, google-dc-optical, mc_flexgrid, mc_icc_overlay, dtb, elastic_optical_networks, b4, yahoo, orchestrating, jetway}. Examples of such transfers include synchronizing search index information \cite{b4}, replication of databases \cite{cassandra, azuresql}, and distribution of high definition videos across CDNs \cite{utube, netflix, jetway, ecoflow, social_inside}. Table \ref{table_0} provides a brief list of how many replicas are made for some applications.

\begin{table}
\small
\begin{center}
\begin{tabular}{ |p{1.5cm}|p{5cm}| }

\hline
\textbf{Service} & \textbf{Replicas} \\
\hline
\hline
Facebook & Across availability regions \cite{rep-facebook}, $\ge 4$ \cite{rep-facebook-2}, for various object types including large machine learning configs \cite{fb-holistic} \\
\hline
CloudBasic SQL Server & Up to $4$ secondary databases with active Geo-Replication (asynchronous) \cite{rep-cloudbasic} \\
\hline
Azure SQL Database & Up to $4$ secondary databases with active Geo-Replication (asynchronous) \cite{rep-azure} \\
\hline
Oracle Directory Server & Up to the number of datacenters owned by an enterprise for regional load balancing of directory servers \cite{rep-oracle-1, rep-oracle-2}\\
\hline
AWS Route $53$ GLB & Across multiple regions and availability zones for global load balancing \cite{route-53} \\
\hline
Youtube & Function of popularity, content potentially pushed to many locations (could be across $\ge33$ datacenters \cite{rep-youtube}) \\
\hline
Netflix & Across $2$ to $4$ availability regions \cite{rep-netflix-regions}, and up to $233$ cache locations \cite{rep-netflix-locations} \\
\hline

\end{tabular}
\end{center}
\caption{Various Services Using Replication} \label{table_0}
\end{table}

In this paper, we focus on transfers that deliver an object from a source to multiple destinations and call them \textbf{Point to Multipoint (P2MP)} transfers. One solution to make such transfers is to initiate multiple independent point-to-point (P2P) transfers that are scheduled separately \cite{mbdt_initial, ssnf, netstitcher, postcard, dtb,  grease, geo_backup_selection, amoeba, tempus, ecoflow, orchestrating, dcroute}. There may however be more efficient ways, in terms of total bandwidth usage and transfer completion times, to perform P2MP transfers by sending at most one copy of the message across any link given that the source datacenter and destination datacenters are known apriori. We present an elegant solution using minimum weight Steiner Trees \cite{steiner_tree_problem} for P2MP transfers that achieves reduced bandwidth usage and tail completion times.

Another approach would be to select trees that connect from sources to all destinations and complete transfers using store-and-forward \cite{mbdt_initial, ssnf, netstitcher, postcard, dtb}. This will impose the overhead of transferring and storing additional copies of objects on intermediate datacenters during the delivery process incurring storage costs as the number of transfers increase, and wasting intra-datacenter bandwidth since such objects have to be stored on a server inside the intermediate datacenters.

Alternatively one can use multicast protocols, which are designed to form group memberships, provide support for members to join or leave, and manage multicast trees as users come and go \cite{ip_multicast, centralized-multicast}. These approaches involve complex management algorithms and protocols to cope with changes in multicast group membership \cite{narada, multicast-challenges} which are unnecessary for our problem given that the participants of each P2MP transfer are fixed.

Application layer multicast techniques \cite{app_layer_multicast, narada} reduce the implementation and management complexity by creating an overlay network across the multicast group. However, since these methods are typically implemented in end-host applications, they lack full visibility to underlying network properties and status, such as topology and available bandwidth, and may still send multiple copies of packets along the same links wasting bandwidth.

\textbf{\textit{How can one efficiently perform P2MP transfers?}} Objective is minimizing tail Transfer Completion Times (TCT) considering limited available bandwidth of networks connecting datacenters and the many transfers that share the network which arrive in an online manner. Also, the system does not have prior knowledge of incoming P2MP transfers; therefore, any solution has to be fast and efficient to deal with them as they arrive.

To perform a P2MP transfer, traffic can be concurrently sent to all destinations over a minimum weight Steiner Tree that connects the source and destinations with transfers' demands as link weights. We refer to such trees as forwarding trees using which we can reduce total bandwidth usage. A central controller with global view of network topology and distribution of load \cite{b4, bwe, swan, tempus, dynamic_pricing, sd_wan} can carefully weigh out various options and select forwarding trees for transfers.

We can implement forwarding trees using SDN \cite{sdn} capable switches that support Group Tables \cite{openflow-1.3.1, of-juniper-explain} such as \cite{of-juniper, of-huawei, of-hp, of-hp-2}. Using this feature, a packet can be replicated and supplied to multiple buckets each processing and forwarding it to a required output switch port. A group might have many buckets, for example one vendor supports up to $32$ \cite{of-juniper-explain}. There is growing vendor support for group tables as later OpenFlow standards are adopted. In this paper, we use abstract simulations to verify our techniques. In the future, we plan on implementing forwarding trees using Group Tables.

\begin{figure}
    \centering
    \includegraphics[width=0.9\columnwidth]{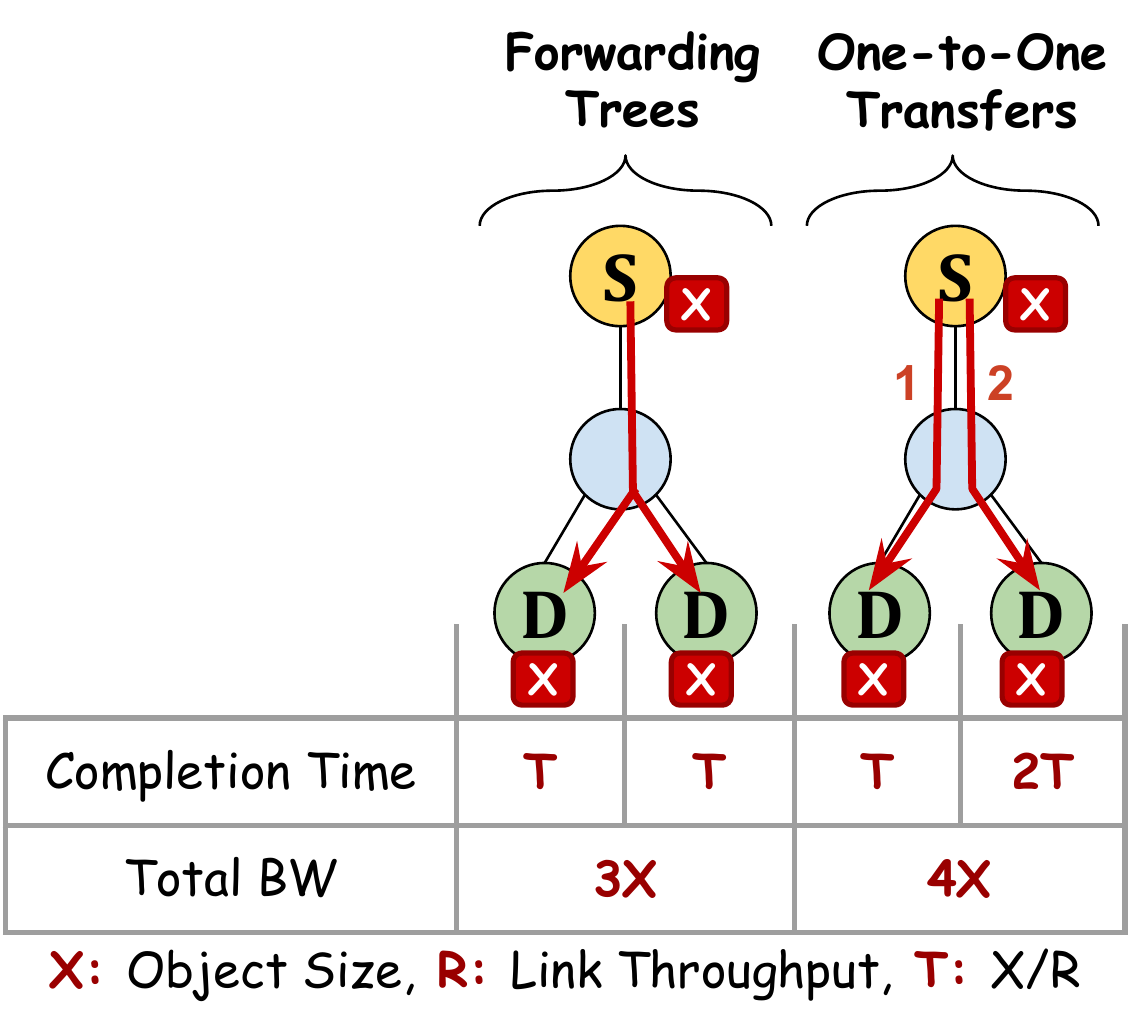}
    \caption{Benefits of using Forwarding Trees}
    \label{fig:p2mp_0}
\end{figure}

\textbf{Motivating Example:} In Figure \ref{fig:p2mp_0}, an object $X$ is to be transferred from datacenter $S$ to two $D$ datacenters considering a link throughput of $R$. In order to send $X$ to destinations, one could initiate individual transfers, but that wastes bandwidth and increases delivery time since the link attached to $S$ turns into a bottleneck.

In this paper, we present an efficient scheme for P2MP transfers called \textbf{DCCast}. It selects forwarding trees according to a weight assignment that tries to balance load across the network. In addition, it uses temporal planning \cite{tempus} and schedules P2MP requests on a first come first serve (FCFS) basis to provide guarantees on completion times. FCFS is shown to reduce tail completion times under light-tailed job size distribution \cite{fcfs-light-tail-opt, caltech-tail} (optimal discipline depends on this distribution \cite{caltech-tail}).

A related concept is Coflows \cite{coflow} that improve performance by jointly scheduling groups of flows with a collective objective. A P2MP transfer can be viewed as a coflow; however unlike other coflows, P2MP transfers present a better opportunity for optimization since the same data is being delivered to different destinations. Using this, DCCast provides the added benefit of reduced bandwidth usage.

We conducted extensive simulation experiments to evaluate DCCast compared with other strategies for picking forwarding trees and scheduling techniques. We performed simulations using synthetic traffic over the Google's GScale topology \cite{b4} with $12$ nodes and $19$ edges and random topologies with $50$ nodes and $150$ edges. Our evaluation metrics include bandwidth usage as well as mean and tail TCT. Our current solution assumes the same class of service for all transfers. In a general setting, transfers may have different priorities and should be allotted resources accordingly (e.g. near real-time video vs. cross-region backups).

We evaluated various forwarding tree selection methods. Clearly, there is benefit in carefully picking trees and we observed up to $43\%$ improvement in completion times while using DCCast compared to random tree selection and up to $29\%$ compared to selection of trees that minimize the maximum load over any edge.

We compared DCCast with P2P schemes by viewing each P2MP transfer as multiple independent transfers and using multipathing ($K$ shortest paths) to spread the load. For GScale topology and with $2$ to $6$ destinations per transfer, DCCast reduced both bandwidth usage and tail TCT by over $20\%$ to $50\%$, respectively, while providing guarantees to users on completion times.

In summary, our contributions are the following:
\vspace{-0.5em}
\begin{itemize}
    \setlength\itemsep{0em}
    \item Prior work based on traffic scheduling and rate allocation \cite{bwe, swan, tempus} uses individual point to point transfers to deliver the same object to multiple places. We improve on this by using forwarding trees.
    \item Prior work on multicasting \cite{ip_multicast, centralized-multicast, narada, app_layer_multicast, multicast-challenges} is focused on managing multicast groups. With apriori knowledge of transfer destinations and demands, DCCast builds forwarding trees that are more efficient. 
    \item DCCast minimizes packet reordering and provides guarantees on completion times.
\end{itemize}

\section{Problem Formulation} \label{formula}
The list of variables and their definitions is provided on table \ref{table_var}. To allow for flexible bandwidth allocation, we consider a slotted timeline \cite{tempus, amoeba, dcroute} where the transmission rate of senders is constant during each timeslot, but can vary from one timeslot to next. This can be achieved via rate-limiting at end-hosts \cite{swan, bwe}. A central scheduler is assumed that receives transfer requests from end-points, calculates their temporal schedule, and informs the end-points of rate-allocations when a timeslot begins. We focus on scheduling large transfers that take more than a few timeslots to finish and therefore, the time to submit a transfer request, calculate the routes, and install forwarding rules is considered negligible in comparison. We assume equal capacity for all links in an online scenario where requests may arrive anytime.

\begin{table}
\small
\begin{center}
\begin{tabular}{ |p{1.2cm}|p{5.5cm}| }

\hline
\textbf{Variable} & \textbf{Definition} \\
\hline
\hline
$R$ & A P2MP transfer \\
\hline
${\cal V}_R$ & Volume of $R$ in bytes \\
\hline
$S_R$ & Source datacenter of $R$ \\
\hline
$\pmb{\mathrm{D_R}}$ & Set$\langle\rangle$ of destinations of $R$ \\
\hline
$G$ & The inter-datacenter network graph \\
\hline
$T$ & A Steiner Tree \cite{steiner_tree_problem} \\
\hline
$\pmb{\mathrm{E_G}}$ & Set$\langle\rangle$ of edges of $G$ \\
\hline
$\pmb{\mathrm{E_T}}$ & Set$\langle\rangle$ of edges of $T$ \\
\hline
$L_e$ & Total load currently scheduled on edge $e$ \\
\hline
$B_e(t)$ & Available bandwidth on edge $e$ at timeslot $t$ \\
\hline
$W$ & Width of a timeslot in seconds \\
\hline

\end{tabular}
\end{center}
\caption{Definition of Variables} \label{table_var}
\end{table}

\section{DCCast} \label{dccast}
\textbf{Forwarding Trees:}
Our proposed approach is, for each P2MP transfer, to jointly route traffic from source to all destinations over a forwarding tree to save bandwidth. Using a single forwarding tree for every transfer also minimizes packet reordering which is known to waste CPU and memory resources at the receiving ends especially at high rates \cite{juggler, mptcphard}. To perform a P2MP transfer $R$ with volume ${\cal V}_R$, the source $S_R$ transmits traffic over a Steiner Tree that spans across $\pmb{\mathrm{D_R}}$. At any timeslot, traffic for any transfer flows with the same rate over all links of a forwarding tree to reach all the destinations at the same time. The problem of scheduling a P2MP transfer then translates to finding a forwarding tree and a transmission schedule over such a tree for every arriving transfer in an online manner. A relevant problem is the Minimum Weight Steiner Tree \cite{steiner_tree_problem} that can help minimize total bandwidth usage with proper weight assignment. Although it is a hard problem, heuristic algorithms exist that often provide near optimal solutions \cite{robins2005tighter, Watel2014}.

\textbf{Scheduling Discipline:}
When forwarding trees are found, we schedule traffic over them according to First Come First Serve (FCFS) policy using all available residual bandwidth on links to minimize the completion times. This allows us to provide guarantees to users on when their transfers will complete upon their arrival. We do not use a preemptive scheme, such as Shortest Remaining Processing Time (SRPT), due to practical concerns: larger transfers might get postponed over and over which might lead to the starvation problem and it is not possible to make promises on exactly when a transfer would complete. Optimal scheduling discipline to minimize tail times rests on transfer size distribution \cite{caltech-tail}.

\SetAlgoVlined
\SetInd{1.2em}{0.5em}
\begin{algorithm}[t]
\KwIn{$R({\cal V}_R,S_R,\pmb{\mathrm{D_R}})$, $G$, $W$, $L_{e}$ and $B_e(t)$ for $e \in \pmb{\mathrm{E_G}}$ and $t > t_{now}$}

\vspace{0.4em}
\KwOut{Forwarding tree (minimum weight Steiner Tree) $T$ and Transmission Schedule of $R$ for $t > t_{now}$}

\vspace{0.4em}
To every edge $e \in \pmb{\mathrm{E_G}}$, assign weight $W_e = (L_{e} + {\cal V}_R)$\;

\vspace{0.4em}
Find the Minimum Weight Steiner Tree $T$ that connects $S_R \cup \pmb{\mathrm{D_R}}$. We used GreedyFLAC \cite{Watel2014, DSTAlgoEvaluation}\;

\vspace{0.4em}
\textit{Schedule $R$ over $T$ to finish as early as possible.}

\begin{adjustwidth}{1em}{0.2em}
$t^{\prime}$ ~$\gets$~ $t_{now}+1$ and ${\cal V}^{\prime}$ ~$\gets$~ ${\cal V}_R$ \;
\While{${\cal V}^{\prime} > 0$} {
 $B_T$ ~$\gets$~ $\pmb{\mathrm{min}}_{e \in \pmb{\mathrm{E_T}}}(B_e(t^{\prime}))$ \;
 Schedule $R$ on $T$ with rate $\pmb{\mathrm{min}}(B_T,\frac{{\cal V}^{\prime}}{W})$ at timeslot $t^{\prime}$ \; 
 $t^{\prime}$ ~$\gets$~ $t^{\prime}+1$ and ${\cal V}^{\prime}$ ~$\gets$~ ${\cal V}^{\prime}-B_T \times W$ \;
}
\end{adjustwidth}

\vspace{0.4em}
\Return{$T$ and the Transmission Schedule of $R$}\;

\caption{Allocate($R$)} \label{algo_1}
\end{algorithm}

\textbf{Algorithms:}
DCCast is made of two algorithms\footnote{Please find an implementation of DCCast algorithms on Github: \url{https://github.com/noormoha/DCCast}}. \textbf{Update()} is executed upon beginning of every timeslot. It simply dispatches the transmission schedule, that is the rate for each transfer, to all senders to adjust their rates via rate-limiting and adjusts $L_{e}~(e \in \pmb{\mathrm{E_G}})$ by deducting the total traffic that was sent over $e$ during current timeslot.

\textbf{Allocate($R$)} is run upon arrival of every request which finds a forwarding tree and schedules $R$ to finish as early as possible. Pseudo-code of this function has been shown in Algorithm \ref{algo_1}. Statically calculating minimal forwarding trees might lead to creation of hot-spots, even if there exists one highly loaded edge that is shared by all of such trees. As a result, DCCast adaptively chooses a forwarding tree that reduces the tail transfer completion times while saving considerable bandwidth.

It is possible that larger trees provide higher available bandwidth by using longer paths through least loaded edges, but using which would consume more overall bandwidth since they send same traffic over more edges. To model this behavior, we use a weight assignment that allows balancing these two possibly conflicting objectives. The weights represent traffic load allocated on links. Selecting links with lower weights will improve load balancing that would be better for future requests. The tradeoff is in avoiding heavier links at the expense of getting larger trees for even distribution of load.

The forwarding tree $T$ selected by Algorithm \ref{algo_1} will have the total weight $\sum_{e \in \pmb{\mathrm{E_T}}}(L_{e} + {\cal V}_R)$. This weight is essentially the total load over $T$ if request $R$ were to be put on it. Selecting trees with minimal total weight will most likely avoid highly loaded edges and larger trees. To find an approximate minimum weight Steiner Tree, we used GreedyFLAC \cite{Watel2014, DSTAlgoEvaluation}, which is quite fast and in practice provides results not far from optimal.

\begin{figure}
    \centering
    \includegraphics[width=\columnwidth]{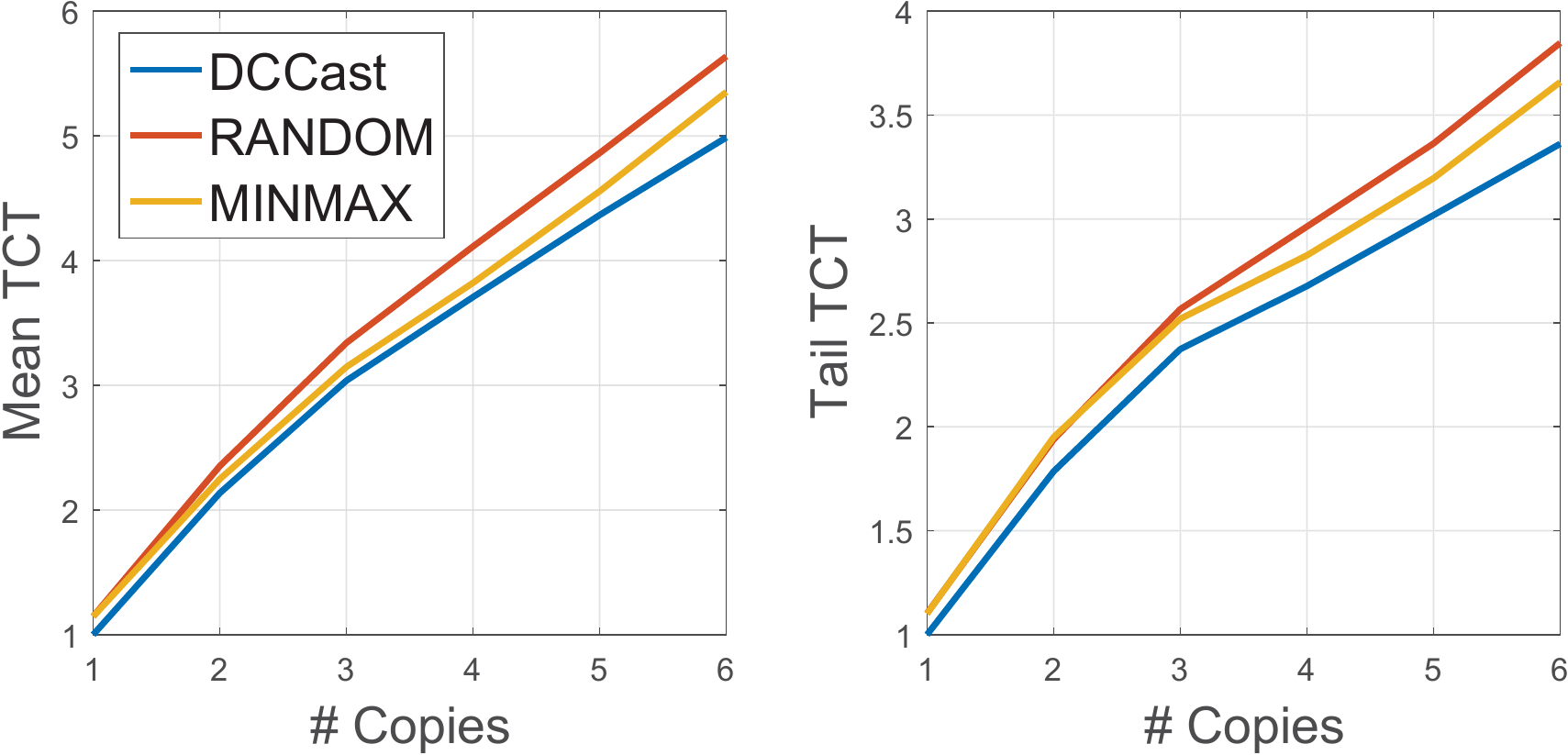}
    \caption{Tree Selection (GScale Topo)}
    \label{fig:p2mp_all_1}
\end{figure}

\begin{figure}
    \centering
    \includegraphics[width=\columnwidth]{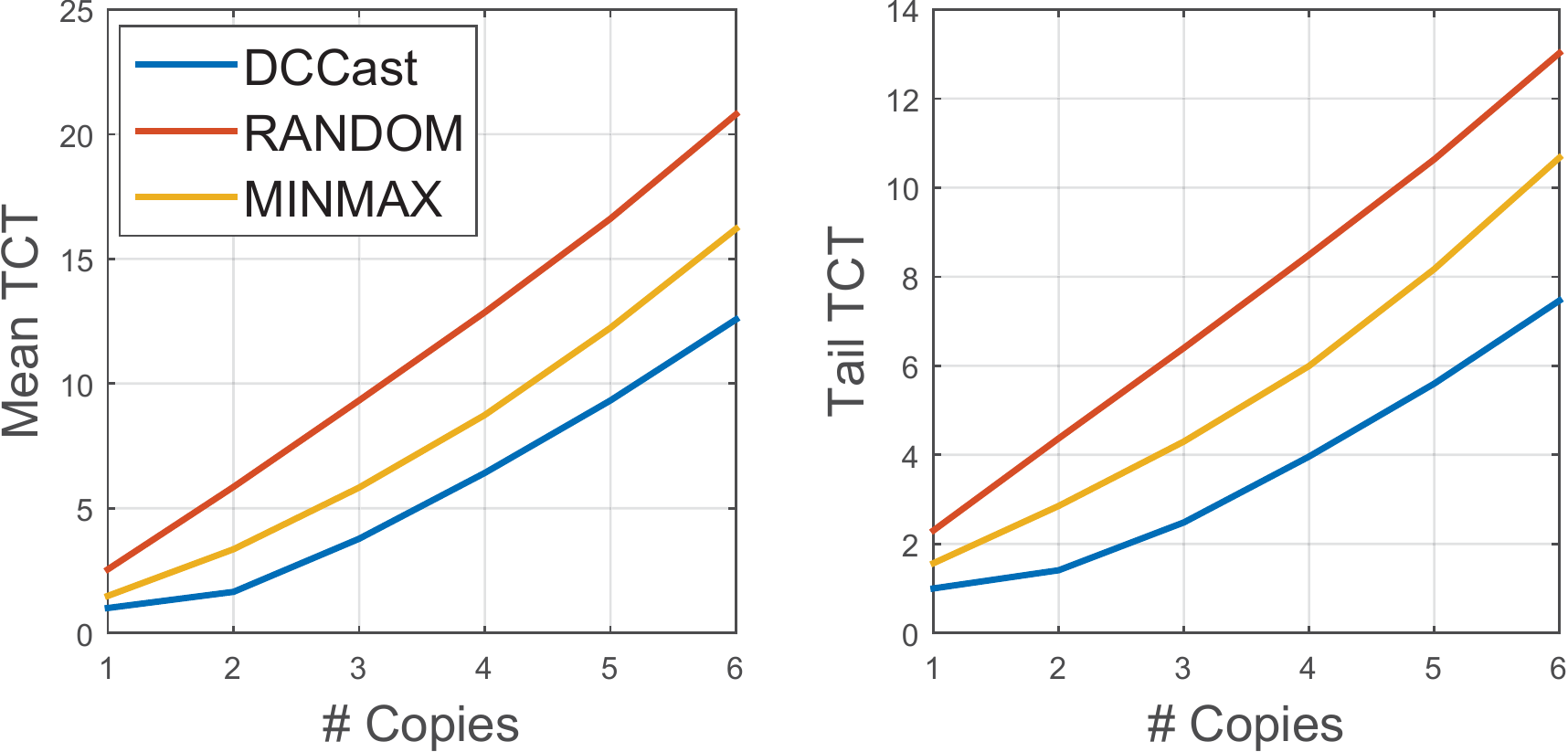}
    \caption{Tree Selection (Random Topo, $\lvert V \lvert=50$)}
    \label{fig:p2mp_all_1_2}
\end{figure}

\section{Evaluation} \label{eval}

\begin{figure}
    \centering
    \includegraphics[width=\columnwidth]{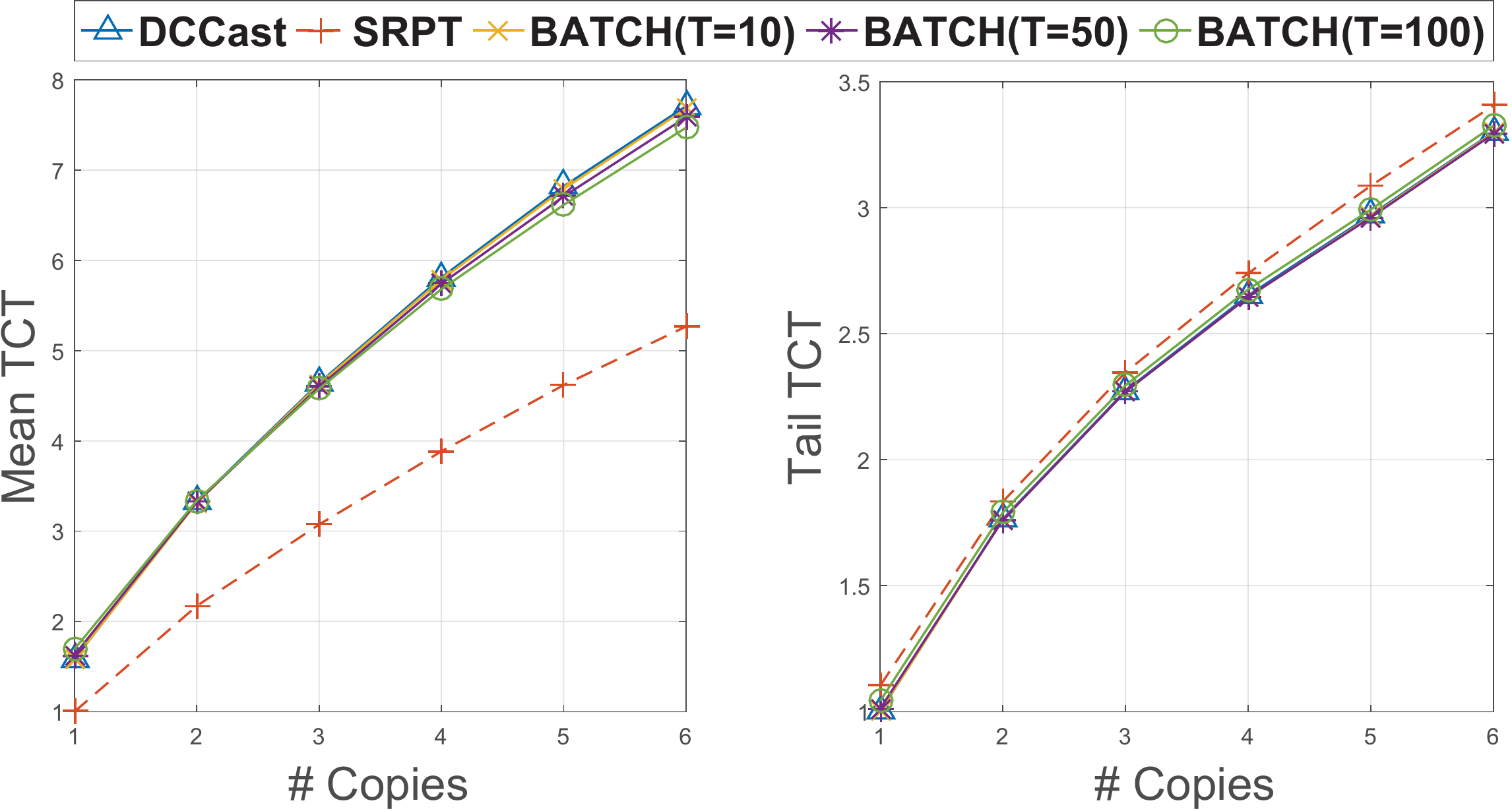}
    \caption{Various Scheduling Policies}
    \label{fig:p2mp_all_2}
\end{figure}

\begin{figure*}
    \centering
    \includegraphics[width=\textwidth]{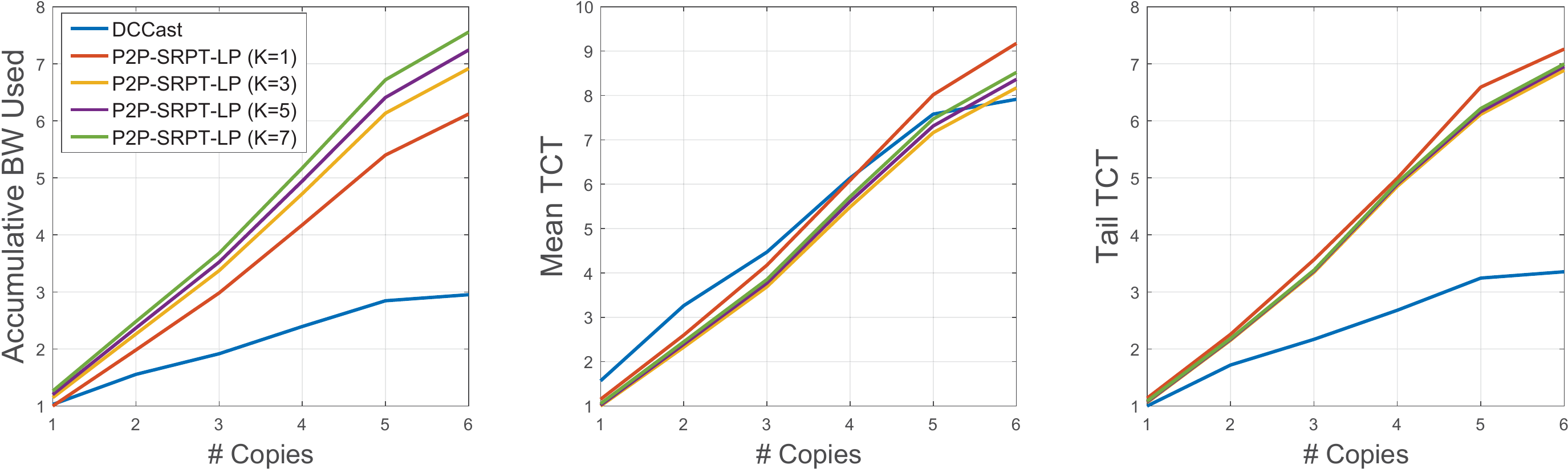}
    \caption{DCCast vs Point-To-Point (P2P-SRPT-LP)}
    \label{fig:p2mp_all_3}
\end{figure*}

\begin{table*}
    \small
    \begin{center}
        \begin{tabular}{ |l|p{13.5cm}| } 
        
\hline
\textbf{Scheme} & \textbf{Method} \\
\hline
\hline
MINMAX      & Selects forwarding trees to minimize maximum load on any link. Schedules traffic using FCFS policy \S \ref{dccast} \\
\hline
RANDOM      & Selects random forwarding trees. Schedules traffic using FCFS policy \S \ref{dccast} \\
\hline
BATCHING    & Batches (enqueues) new requests arriving in time windows of $T$. At the end of batching windows, jointly schedules all new requests according to Shortest Job First (SJF) policy and picks their forwarding trees using weight assignment of Algorithm \ref{algo_1} \\
\hline
SRPT        & Upon arrival of a new request, jointly reschedules all existing requests and the new request according to SRPT policy \S \ref{dccast} and picks new forwarding trees for all requests using weight assignment of Algorithm \ref{algo_1} \\
\hline
P2P-SRPT-LP & Views each P2MP request as multiple independent point-to-point (P2P) requests. Uses a Linear Programming (LP) model along with SRPT policy \S \ref{dccast} to (re)schedule each request over $K$-Shortest Paths between its source and destination upon arrival of new requests \\
\hline
P2P-FCFS-LP & Similar to above while using FCFS policy \S \ref{dccast} \\
\hline
            
        \end{tabular}
    \end{center}
\caption{Schemes Used for Comparison}
\label{table_1}
\end{table*}

We evaluated DCCast using synthetic traffic. We assumed a total capacity of $1.0$ for each timeslot over every link. The arrival of requests followed a Poisson distribution with rate $\lambda_{P2MP} = 1$. Demand of every request $R$ was calculated using an exponential distribution with mean $20$ added to a constant value of $10$ (fixing the minimum demand to $10$). All simulations were performed over as many timeslots as needed to finish all requests with arrival time of last request set to be $500$ or less. Presented results are normalized by minimum values in each chart.

We measure three different metrics: \textbf{total bandwidth used} and \textbf{mean and tail TCT}. The total bandwidth used is the sum of all traffic over all timeslots and all links. The completion time of a transfer is defined as its arrival time to the time its last bit is delivered to the destination(s). We performed simulations using Google's GScale topology \cite{b4}, with $12$ nodes and $19$ edges, on a single machine (Intel Core i7-6700T and 24 GBs of RAM). All simulations were coded in Java and used Gurobi Optimizer \cite{gurobi} to solve linear programs for P2P schemes. We increased the destinations (copies) for each object from $1$ to $6$ picking recipients according to uniform distribution. Table \ref{table_1} shows list of considered schemes (first $4$ are P2MP schemes and last $2$ are P2P).

We tried various forwarding tree selection criteria over both GScale topology and a larger random topology with $50$ nodes and $150$ edges as shown in Figures \ref{fig:p2mp_all_1} and \ref{fig:p2mp_all_1_2}, respectively. In case of GScale, DCCast performs slightly better than RANDOM and MINMAX in completion times while using equal overall bandwidth (not in figure). In case of larger random topologies, DCCast's dominance is more obvious regarding completion times while using same or less bandwidth (not in figure).

We also experimented various scheduling disciplines over forwarding trees as shown in Figure \ref{fig:p2mp_all_2}. The SRPT discipline performs considerably better regarding mean completion times; it however may lead to starvation of larger transfers if smaller ones keep arriving. It has to compute and install new forwarding trees and recalculate the whole schedule, for all requests currently in the system with residual demands, upon arrival of every new request. This could impose significant rule installation overhead which is considered negligible in our evaluations. It might also lead to lots of packet loss and reordering. Batching improves performance marginally compared to DCCast and could be an alternate road to take. Generally, a smaller batch size results in a smaller initial scheduling latency while a larger batch size makes it possible to employ collective knowledge of many requests in a batch for optimized scheduling. Batching might be more effective for systems with bursty request arrival patterns. All schemes performed almost similarly regarding tail completion times and total bandwidth usage (not in figure).

In Figure \ref{fig:p2mp_all_3} we compare DCCast with a Point-To-Point scheme (P2P-SRPT-LP) using SRPT scheduling policy which uses various number of shortest paths ($K$) and delivers each copy independently. The total bandwidth usage is close for all schemes when there is only one destination per request. Both bandwidth usage and tail completion times of DCCast are up to $50\%$ less than that of P2P-SRPT-LP as the number of destinations per transfer increases. Although DCCast follows the FCFS policy, its mean completion time is close to that of P2P-SRPT-LP and surpasses it for $6$ copies due to bandwidth savings which leave more headroom for new transfers.

In a different experiment, we compared DCCast with P2P-FCFS-LP. DCCast again saved up to $50\%$ bandwidth and reduced tail completion times by up to almost $50\%$ while increasing the number of destinations per transfer.

\textbf{Computational Overhead:} We used a network with $50$ nodes and $300$ edges and considered P2MP transfers with $5$ destinations per transfer. Transfers were generated according to Poisson distribution with arrival times ranging from $0$ to $1000$ timeslots and the simulation ran until all transfers were completed. Mean processing time of a single \textit{timeslot} increased from $1.2ms$ to $50ms$ per timeslot while increasing $\lambda_{P2MP}$ from $1$ to $10$. Mean processing time of a single \textit{transfer} (which accounts for finding a tree and scheduling the transfer) was $1.2ms$ and $5ms$ per transfer for $\lambda_{P2MP}$ equal to $1$ and $10$, respectively. This is negligible compared to timeslot lengths of minutes in prior work \cite{amoeba}.

\section{Conclusions and Future Work}
In this paper, we presented DCCast, which aims to reduce the total network bandwidth usage while minimizing transfer completion times using point to multipoint (P2MP) transfers. To save bandwidth, DCCast uses forwarding trees that connect source of a P2MP transfer to its destinations. Selection of forwarding trees is performed in a way that attempts to balance load across the network by avoiding highly loaded links while reducing bandwidth usage by choosing smaller trees. To provide guarantees on transfer completion times, DCCast schedules new traffic to finish as early as possible while not changing the schedule of already allocated transfers. Our evaluations show that DCCast can significantly reduce bandwidth usage compared to viewing each P2MP transfer as multiple independent transfers.

In the future, we would like to perform testbed experiments using traces of real traffic. An alternate scheduling scheme to what we proposed would be Fair Sharing which we aim to study. Next, we would like to evaluate DCCast using a mix of P2MP transfers with different number of destinations to better understand how various applications might interact. It would be interesting to consider multiple classes of traffic with different priorities as well. Moreover, further investigation is needed to measure the fraction of traffic with multiple destinations which benefit from forwarding trees. Finally, it is necessary to study approaches for handling and recovering from failures.

\section{Acknowledgments}
We would like to thank our shepherd Sindhu Ghanta and the anonymous reviewers of the HotCloud workshop whose comments helped us greatly improve the quality of this paper.

\section{Discussion Topics}
Prior work on inter-datacenter transfers uses a combination of various techniques, such as rate-control \cite{bwe, swan}, temporal planning \cite{tempus}, store-and-forward \cite{dtb}, and topology changes \cite{owan}, to improve the performance and efficiency of point-to-point (P2P) inter-datacenter transfers. In this paper, we focused on point-to-multipoint (P2MP) transfers and a new direction in which forwarding trees are used to deliver an object simultaneously to all destinations. We discussed the benefits of this approach and presented evaluations to back up our proposal.

The extent to which our approach can benefit operators as well as cloud and datacenter applications used in industry would be one possible discussion point. This depends partly on the type of applications, desired level of reliability, and customer SLOs such as user access latency. Finally, a crucial factor is the fraction of transfers with multiple destinations, which are ones that can actually benefit from use of forwarding trees.

There are several discussion points regarding implementation. In case of using tools, such as SDN, to cheaply setup and tear down forwarding trees, how can we meet the necessary performance metrics, such as delay, considering number of required forwarding rules, transfer arrival rate, and latency overhead of rule installations? What are possible tensions between performance efficiency and practical feasibility of techniques?

Another topic of discussion would be selection of the forwarding trees. We presented and evaluated three possible selection methods. However selection of forwarding trees that reduce usage of bandwidth while minimizing transfer completion times is an open problem. In addition, would it be better to select and setup trees dynamically, or statically build many trees and use them?

Scalability of our approach considering various network topologies, different network sizes as well as arrival rate of transfers, effectiveness of the scheduling discipline used in satisfying users and operators, and handling of network or end-point failures would be other possible discussion topics. Would it be possible to apply known methods of improving scalability if necessary?

{\footnotesize \bibliographystyle{unsrt}
\bibliography{citations.bib}}

\begin{thebibliography}{10}

\bibitem{azure}
Microsoft azure: Cloud computing platform \& services.
\newblock \url{https://azure.microsoft.com/}.

\bibitem{google}
Compute engine - iaas - google cloud platform.
\newblock \url{https://cloud.google.com/compute/}.

\bibitem{aws}
Amazon web services (aws) - cloud computing services.
\newblock \url{https://aws.amazon.com/}.

\bibitem{orchestrating}
Yu~Wu, Zhizhong Zhang, Chuan Wu, et~al.
\newblock Orchestrating bulk data transfers across geo-distributed datacenters.
\newblock {\em IEEE Transactions on Cloud Computing}, PP(99):1--1, 2015.

\bibitem{tempus}
Srikanth Kandula, Ishai Menache, Roy Schwartz, et~al.
\newblock Calendaring for wide area networks.
\newblock {\em ACM SIGCOMM}, 44(4):515--526, 2015.

\bibitem{swan}
Chi-Yao Hong, Srikanth Kandula, Ratul Mahajan, et~al.
\newblock Achieving high utilization with software-driven wan.
\newblock In {\em ACM SIGCOMM}, pages 15--26, 2013.

\bibitem{b4}
Sushant Jain, Alok Kumar, Subhasree Mandal, et~al.
\newblock B4: Experience with a globally-deployed software defined wan.
\newblock {\em ACM SIGCOMM}, 43(4):3--14, 2013.

\bibitem{mesa}
Ashish Gupta, Fan Yang, Jason Govig, et~al.
\newblock Mesa: A geo-replicated online data warehouse for google's advertising
  system.
\newblock {\em Communications of the ACM}, 59(7):117--125, June 2016.

\bibitem{mdcc}
Tim Kraska, Gene Pang, Michael~J Franklin, Samuel Madden, and Alan Fekete.
\newblock Mdcc: Multi-data center consistency.
\newblock In {\em ACM Eurosys}, pages 113--126, 2013.

\bibitem{owan}
Xin Jin, Yiran Li, Da~Wei, Siming Li, et~al.
\newblock Optimizing bulk transfers with software-defined optical wan.
\newblock In {\em ACM SIGCOMM}, pages 87--100, 2016.

\bibitem{google-dc-optical}
Xiaoxue Zhao, Vijay Vusirikala, Bikash Koley, et~al.
\newblock The prospect of inter-data-center optical networks.
\newblock {\em IEEE Communications Magazine}, 51(9):32--38, 2013.

\bibitem{mc_flexgrid}
Ll~Gifre, F~Paolucci, J~Marhuenda, et~al.
\newblock Experimental assessment of inter-datacenter multicast connectivity
  for ethernet services in flexgrid networks.
\newblock In {\em Proceedings of European Conference on Optical
  Communications}, pages 1--3, 2014.

\bibitem{mc_icc_overlay}
Yasuhiro Miyao.
\newblock An overlay architecture of global inter-data center networking for
  fast content delivery.
\newblock In {\em IEEE International Conference on Communications (ICC)}, pages
  1--6, 2011.

\bibitem{dtb}
Nikolaos Laoutaris, Georgios Smaragdakis, Rade Stanojevic, et~al.
\newblock Delay-tolerant bulk data transfers on the internet.
\newblock {\em IEEE/ACM Transactions on Networking}, 21(6):1852--1865, December
  2013.

\bibitem{elastic_optical_networks}
Ping Lu, Liang Zhang, Xiahe Liu, et~al.
\newblock Highly efficient data migration and backup for big data applications
  in elastic optical inter-data-center networks.
\newblock {\em IEEE Network}, 29(5):36--42, 2015.

\bibitem{yahoo}
Yingying Chen, Sourabh Jain, Vijay~Kumar Adhikari, et~al.
\newblock A first look at inter-data center traffic characteristics via yahoo!
  datasets.
\newblock In {\em IEEE INFOCOM}, pages 1620--1628, 2011.

\bibitem{jetway}
Yuan Feng, Baochun Li, and Bo~Li.
\newblock Jetway: Minimizing costs on inter-datacenter video traffic.
\newblock In {\em ACM International conference on Multimedia}, pages 259--268,
  2012.

\bibitem{cassandra}
Multi-datacenter replication in cassandra.
\newblock \url{http://www.datastax.com/dev/blog/multi-datacenter-replication},
  2012.

\bibitem{azuresql}
Azure sql database now supports powerful geo-replication features for all
  service tiers.
\newblock
  \url{https://azure.microsoft.com/en-us/blog/azure-sql-database-now-supports-powerful-geo-replication-features-on-all-service-tiers/},
  2016.

\bibitem{utube}
Vijay~Kumar Adhikari, Sourabh Jain, Yingying Chen, and Zhi-Li Zhang.
\newblock Vivisecting youtube: An active measurement study.
\newblock In {\em IEEE INFOCOM}, pages 2521--2525, 2012.

\bibitem{netflix}
{Meshenberg, Ruslan and Gopalani, Naresh and Kosewski, Luke}.
\newblock {Active-Active for Multi-Regional Resiliency}.
\newblock
  \url{http://techblog.netflix.com/2013/12/active-active-for-multi-regional.html},
  2013.

\bibitem{ecoflow}
Yuhua Lin, Haiying Shen, and Liuhua Chen.
\newblock Ecoflow: An economical and deadline-driven inter-datacenter video
  flow scheduling system.
\newblock In {\em Proceedings of the ACM international conference on
  Multimedia}, pages 1059--1062, 2015.

\bibitem{social_inside}
Arjun Roy, Hongyi Zeng, Jasmeet Bagga, George Porter, and Alex~C. Snoeren.
\newblock Inside the social network's (datacenter) network.
\newblock In {\em ACM SIGCOMM}, pages 123--137, 2015.

\bibitem{rep-facebook}
Companies like facebook and google have multiple data centers. do these
  datacenters all store copies of the same information?
\newblock
  \url{https://www.quora.com/Companies-like-Facebook-and-Google-have-multiple-data-centers-Do-these-datacenters-all-store-copies-of-the-same-information}.
\newblock visited on March 3, 2017.

\bibitem{rep-facebook-2}
Where are the facebook servers located worldwide?
\newblock
  \url{https://www.quora.com/Where-are-the-Facebook-servers-located-worldwide}.
\newblock visited on March 3, 2017.

\bibitem{fb-holistic}
Chunqiang Tang, Thawan Kooburat, Pradeep Venkatachalam, et~al.
\newblock Holistic configuration management at facebook.
\newblock In {\em Proceedings of the Symposium on Operating Systems
  Principles}, pages 328--343. ACM, 2015.

\bibitem{rep-cloudbasic}
Geo-replication/multi-ar (active).
\newblock \url{http://cloudbasic.net/documentation/geo-replication-active/}.
\newblock visited on March 5, 2017.

\bibitem{rep-azure}
Overview: Sql database active geo-replication.
\newblock
  \url{https://docs.microsoft.com/en-us/azure/sql-database/sql-database-geo-replication-overview}.
\newblock visited on March 5, 2017.

\bibitem{rep-oracle-1}
Using replication across multiple data centers.
\newblock
  \url{https://docs.oracle.com/cd/E20295_01/html/821-1217/fpcoo.html\#aalgm}.
\newblock visited on March 11, 2017.

\bibitem{rep-oracle-2}
Understanding oracle internet directory replication.
\newblock
  \url{https://docs.oracle.com/cd/E28280_01/admin.1111/e10029/oid_replic.htm\#OIDAG2201}.
\newblock visited on March 11, 2017.

\bibitem{route-53}
Global load balancing using aws route 53.
\newblock
  \url{https://www.sumologic.com/blog-amazon-web-services/aws-route-53-global-load-balancing/}.
\newblock visited on March 14, 2017.

\bibitem{rep-youtube}
Ruben Torres, Alessandro Finamore, Jin~Ryong Kim, Marco Mellia, Maurizio~M
  Munafo, and Sanjay Rao.
\newblock Dissecting video server selection strategies in the youtube cdn.
\newblock In {\em Distributed Computing Systems (ICDCS)}, pages 248--257. IEEE,
  2011.

\bibitem{rep-netflix-regions}
How netflix leverages multiple regions to increase availability (arc305).
\newblock \url{https://www.slideshare.net/AmazonWebServices/arc305-28387146}.
\newblock visited on March 3, 2017.

\bibitem{rep-netflix-locations}
Mapping netflix: Content delivery network spans 233 sites.
\newblock
  \url{http://datacenterfrontier.com/mapping-netflix-content-delivery-network/}.
\newblock visited on March 3, 2017.

\bibitem{mbdt_initial}
Yiwen Wang, Sen Su, Alex~X Liu, et~al.
\newblock Multiple bulk data transfers scheduling among datacenters.
\newblock {\em Computer Networks}, 68:123--137, 2014.

\bibitem{ssnf}
Yang Yu, Wang Rong, and Wang Zhijun.
\newblock Ssnf: Shared datacenter mechanism for inter-datacenter bulk transfer.
\newblock In {\em IEEE International Conference on Advanced Cloud and Big Data
  (CBD)}, pages 184--189, 2014.

\bibitem{netstitcher}
Nikolaos Laoutaris, Michael Sirivianos, Xiaoyuan Yang, and Pablo Rodriguez.
\newblock Inter-datacenter bulk transfers with netstitcher.
\newblock In {\em ACM SIGCOMM}, pages 74--85, 2011.

\bibitem{postcard}
Yuan Feng, Baochun Li, and Bo~Li.
\newblock Postcard: Minimizing costs on inter-datacenter traffic with
  store-and-forward.
\newblock In {\em IEEE International Conference on Distributed Computing
  Systems Workshops}, pages 43--50, 2012.

\bibitem{grease}
Thyaga Nandagopal and Krishna~PN Puttaswamy.
\newblock Lowering inter-datacenter bandwidth costs via bulk data scheduling.
\newblock In {\em IEEE International Symposium on Cluster, Cloud and Grid
  Computing (CCGrid)}, pages 244--251, 2012.

\bibitem{geo_backup_selection}
Jingjing Yao, Ping Lu, Long Gong, and Zuqing Zhu.
\newblock On fast and coordinated data backup in geo-distributed optical
  inter-datacenter networks.
\newblock {\em IEEE Journal of Lightwave Technology}, 33(14):3005--3015, 2015.

\bibitem{amoeba}
Hong Zhang, Kai Chen, Wei Bai, et~al.
\newblock Guaranteeing deadlines for inter-datacenter transfers.
\newblock In {\em ACM Eurosys}, page~20, 2015.

\bibitem{dcroute}
Mohammad Noormohammadpour, Cauligi~S Raghavendra, and Sriram Rao.
\newblock Dcroute: Speeding up inter-datacenter traffic allocation while
  guaranteeing deadlines.
\newblock In {\em High Performance Computing, Data, and Analytics (HiPC)}.
  IEEE, 2016.

\bibitem{steiner_tree_problem}
FK~Hwang and Dana~S Richards.
\newblock Steiner tree problems.
\newblock {\em Networks}, 22(1):55--89, 1992.

\bibitem{ip_multicast}
Michelle Cotton, Leo Vegoda, and David Meyer.
\newblock Iana guidelines for ipv4 multicast address assignments, 2010.

\bibitem{centralized-multicast}
Srinivasan Keshav and Sanjoy Paul.
\newblock Centralized multicast.
\newblock In {\em IEEE International Conference on Network Protocols}, pages
  59--68, 1999.

\bibitem{narada}
Yang-hua Chu, Sanjay~G Rao, Srinivasan Seshan, and Hui Zhang.
\newblock A case for end system multicast.
\newblock {\em IEEE Journal on selected areas in communications},
  20(8):1456--1471, 2002.

\bibitem{multicast-challenges}
Bob Quinn and Kevin Almeroth.
\newblock Ip multicast applications: Challenges and solutions, 2001.

\bibitem{app_layer_multicast}
Suman Banerjee, Bobby Bhattacharjee, and Christopher Kommareddy.
\newblock Scalable application layer multicast.
\newblock In {\em Proceedings of the ACM Conference on Applications,
  Technologies, Architectures, and Protocols for Computer Communications},
  pages 205--217, 2002.

\bibitem{bwe}
Alok Kumar, Sushant Jain, Uday Naik, Anand Raghuraman, et~al.
\newblock Bwe: Flexible, hierarchical bandwidth allocation for wan distributed
  computing.
\newblock In {\em ACM SIGCOMM}, pages 1--14, 2015.

\bibitem{dynamic_pricing}
Virajith Jalaparti, Ivan Bliznets, Srikanth Kandula, Brendan Lucier, and Ishai
  Menache.
\newblock Dynamic pricing and traffic engineering for timely inter-datacenter
  transfers.
\newblock In {\em ACM SIGCOMM}, pages 73--86, 2016.

\bibitem{sd_wan}
Xin Jin, Yiran Li, Da~Wei, Siming Li, et~al.
\newblock Optimizing bulk transfers with software-defined optical wan.
\newblock In {\em ACM SIGCOMM}, pages 87--100, 2016.

\bibitem{sdn}
Nick McKeown.
\newblock Software-defined networking.
\newblock {\em IEEE INFOCOM keynote talk}, 17(2):30--32, 2009.

\bibitem{openflow-1.3.1}
Ben Pfaff, Bob Lantz, Brandon Heller, et~al.
\newblock Openflow switch specification, version 1.3.1 (wire protocol 0x04).
\newblock
  \url{https://www.opennetworking.org/images/stories/downloads/sdn-resources/onf-specifications/openflow/openflow-spec-v1.3.1.pdf},
  2012.

\bibitem{of-juniper-explain}
Understanding how the openflow group action works.
\newblock
  \url{https://www.juniper.net/documentation/en_US/junos/topics/concept/junos-sdn-openflow-groups.html}.
\newblock visited on March 14, 2017.

\bibitem{of-juniper}
Openflow v1.3.1 compliance matrix for devices running junos os.
\newblock
  \url{https://www.juniper.net/documentation/en_US/junos/topics/reference/general/junos-sdn-openflow-v1.3.1-compliance-matrix.html}.
\newblock visited on March 14, 2017.

\bibitem{of-huawei}
S12700 series agile switches.
\newblock
  \url{http://e.huawei.com/us/products/enterprise-networking/switches/campus-switches/s12700}.
\newblock visited on March 14, 2017.

\bibitem{of-hp}
Hp 5920 \& 5900 switch series openflow command reference.
\newblock
  \url{http://h20565.www2.hpe.com/hpsc/doc/public/display?sp4ts.oid=5221896&docId=emr_na-c04089449&docLocale=en_US}.
\newblock visited on March 14, 2017.

\bibitem{of-hp-2}
Hp 5130 ei switch series openflow configuration guide.
\newblock
  \url{http://h20565.www2.hpe.com/hpsc/doc/public/display?sp4ts.oid=7399420&docLocale=en_US&docId=emr_na-c04771714}.
\newblock visited on March 14, 2017.

\bibitem{fcfs-light-tail-opt}
Alexander~L Stolyar and Kavita Ramanan.
\newblock Largest weighted delay first scheduling: Large deviations and
  optimality.
\newblock {\em Annals of Applied Probability}, pages 1--48, 2001.

\bibitem{caltech-tail}
Adam Wierman and Bert Zwart.
\newblock Is tail-optimal scheduling possible?
\newblock {\em Operations Research}, 60(5):1249--1257, 2012.

\bibitem{coflow}
Mosharaf Chowdhury and Ion Stoica.
\newblock Coflow: A networking abstraction for cluster applications.
\newblock In {\em ACM HotNets}, pages 31--36, 2012.

\bibitem{juggler}
Yilong Geng, Vimalkumar Jeyakumar, Abdul Kabbani, and Mohammad Alizadeh.
\newblock Juggler: a practical reordering resilient network stack for
  datacenters.
\newblock In {\em ACM Eurosys}, page~20, 2016.

\bibitem{mptcphard}
Costin Raiciu, Christoph Paasch, Sebastien Barre, et~al.
\newblock How hard can it be? designing and implementing a deployable multipath
  tcp.
\newblock In {\em NSDI}, pages 29--29, 2012.

\bibitem{robins2005tighter}
Gabriel Robins and Alexander Zelikovsky.
\newblock Tighter bounds for graph steiner tree approximation.
\newblock {\em Journal on Discrete Mathematics}, 19(1):122--134, 2005.

\bibitem{Watel2014}
Dimitri Watel and Marc-Antoine Weisser.
\newblock {\em A Practical Greedy Approximation for the Directed Steiner Tree
  Problem}, pages 200--215.
\newblock Springer International Publishing, 2014.

\bibitem{DSTAlgoEvaluation}
Evaluation of approximation algorithms for the directed steiner tree problem.
\newblock \url{https://github.com/mouton5000/DSTAlgoEvaluation}.
\newblock visited on Apr 27, 2017.

\bibitem{gurobi}
Inc. Gurobi~Optimization.
\newblock Gurobi optimizer reference manual.
\newblock \url{http://www.gurobi.com}, 2016.

\end{thebibliography}

\end{document}